\documentclass[twocolumn,aps,prd,nofootinbib,superscriptaddress]{revtex4-1}
\usepackage[colorlinks=false,linkcolor=black,citecolor=black]{hyperref}
\usepackage{graphicx}
\usepackage{subfig}
\usepackage{amstext,amsmath,amsfonts,amssymb,amsthm}
\usepackage[format=plain,justification=centerlast]{caption}
\usepackage{hyperref} 
\usepackage{color}

\usepackage{verbatim}
\def\app#1{{Appendix~\ref{#1}}}
\def\sec#1{{Section~\ref{#1}}}
\def\eq#1{{Eq.~(\ref{#1})}}

\def\fig#1{{Fig.~\ref{#1}}}

\def\dd{\mathrm{d}}
\def\be{\begin{equation}}
\def\ee{\end{equation}}
\def\bes{\begin{eqnarray}}
\def\ees{\end{eqnarray}}
\def\ba{\begin{align}}
\def\ea{\end{align}}
\def\bwt{\begin{widetext}}
\def\ewt{\end{widetext}}

\def\f{\frac}

\def\nn{\nonumber}

\begin{document}

\title{Black hole kinematics: the ``in"-vacuum energy density and flux for different observers}
\date{\today}
\author{Suprit Singh}
\email{suprit@iucaa.ernet.in}
\affiliation{IUCAA, Ganeshkhind, Pune 411007, INDIA.}
\author{Sumanta Chakraborty}
\email{sumanta@iucaa.ernet.in}
\affiliation{IUCAA, Ganeshkhind, Pune 411007, INDIA.}

\begin{abstract}
We have investigated the local invariant scalar observables - energy density and flux - which explicitly depend on the kinematics of the concerned observers in the thin null shell gravitational collapse geometry. The use of globally defined null coordinates allows for the definition of a unique in-vacuum for the scalar field propagating in this background. Computing the stress-energy tensor for this scalar field, we work out the energy density and flux for the static observers outside the horizon and then consider the radially in-falling observers who fall in from some specified initial radius all the way through the horizon and inside to the eventual singularity. Our results confirm the thermal Tolman-shifted energy density and fluxes for the static observers which diverge at the horizon. For the in-falling observer starting from far off, both the quantities -- energy density and flux at the horizon crossing are \emph{regular and finite}. For example, the flux at the horizon for the in-falling observer from infinity is approximately 24 times the flux for the observer at infinity. Compared with the static observers in the near-horizon region, this is quite small. Both the quantities grow as the in-fall progresses inside the horizon and diverge at the singularity.     
\end{abstract}

\maketitle

\section{Introduction}

In classical general relativity, the geometry of spacetime is sourced by the matter stress-energy tensor taken to be classical. Since matter is inherently quantum mechanical, we need to consistently combine the ideas of general relativity with quantum mechanics of matter fields. In the absence of a viable theory of quantum gravity, the standard approach is to consider the semi-classical modification of the field equations as $G_{ab} = \kappa\, (T_{ab} + \langle \hat{T}_{ab}\rangle_{\mathrm{ren}})$ such that the background spacetime is treated classically, sourced by the background stress-energy tensor, $T_{ab}$, while all other fields propagating on this background are quantized, backreacting on the background through the renormalized expectation value of the field stress energy tensor $\langle \hat{T}_{ab}\rangle_{\mathrm{ren}}$. This permits us to do quantum field theory in curved spacetime and has given rise to some very interesting results in the case of black holes, cosmological spacetimes and so on (see for details \cite{birrel,Helfer:2003,Fabbri:2005, Mukhanov:2007, Parker:2009}). It is important to emphasize here that in the semi-classical approximation, we ignore the effect of fluctuations of $\hat{T}_{ab}$ and treat it as a classical quantity which enters the field equations, back-reacting on the geometry. While considering the back-reaction, we of course, need to determine the fluctuation in $\hat{T}_{ab}$ which would then test the validity of the semi-classical approximation.

There are two ways in which one can describe the physical content of the vacuum of a quantum field in a gravitational background. The first is using model of a particle detector \cite{Unruh76} which clicks (makes a transition in its internal energy levels) whenever it encounters an excitation of the field. The response function of a detector is essentially given by the Fourier transform of the two-point function of the field with respect to proper time on the detector's trajectory. This very definition has a slight drawback that the Fourier transform with respect to different time coordinates can give different results. Also, the response function depends on the complete history along the detector's trajectory and hence on the global structure of the spacetime. The other approach, free of these pathologies, is constructing two local observables (see \cite{ford1993, ford1995, ford1996,Visser:1996a,Visser:1996b,Visser:1996c}) from the renormalized stress-energy tensor, an invariant quantity and a measure of the ``vacuum activity". Given the $4$-velocity, $u^{a}$ of an observer and a normal $n_{a}$ such that $n_au^a = 0$, we have,
\be
U=\langle \hat{T}_{ab}\rangle u^{a}u^{b}\,;\hspace{10pt}F=-\langle \hat{T}_{ab}\rangle u^{a}n^{b}
\ee
as the energy density and flux in the frame of that observer. Both energy density and flux are local invariant scalars and have an explicit dependence on the kinematics of the observers. The next thing that one can do, to take into account the fluctuations in $\hat{T}_{ab}$, is to consider a detector coupled to the stress-energy tensor of the field \cite{Padmanabhan:1987} which responds to the two-point function of the stress-energy tensor. We shall, however, work within the semi-classical approximation in this paper and neglect the fluctuations of $\hat{T}_{ab}$ in this limit. 

\begin{figure}[t!]
\centering
\includegraphics[scale=1.5]{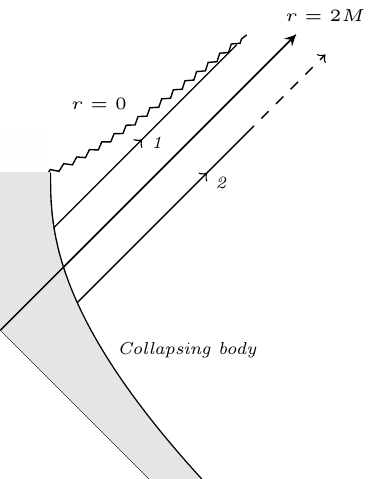}
\caption{The Kruskal diagram of the gravitationally collapsing body showing two rays near the horizon - inside and outside. While the outside one goes off to asymptotic infinity and gives the standard Hawking effect, the inside one hits the singularity in a finite time. As the outside ray gives rise to the standard Hawking effect at infinity, one can ask about the effects of the inside ray.}
\label{Kruskaldiag}
\end{figure}

It is well-known (see \cite{Hawking74,Hawking75}) that the asymptotic observers see a thermal flux with temperature $T_H=1/8\pi M$ being radiated from the black hole horizon. However, we can have non-asymptotic observers and the energy density and flux will be different for each of them depending on their trajectories. There have been many attempts at answering the question, ``What do these non-asymptotic observers see?". The responses to this question include the study of particle detectors on various non-asymptotic trajectories giving the ``effective" temperatures \cite{Barbado11,ms2013} measured on those trajectories. For example, the effective temperature perceived near the horizon for a radially in-falling observer from infinity is $4\,T_H$ and the response is not thermal because of non-stationarity. This is also confirmed \cite{ms2013} by computing the flux at horizon crossing for in-falling observers. However, these answers stop at the horizon and one would like to know what happens on the inside too. Even though the observers who enter the trapped, non-stationary region are doomed to hit the eventual singularity, the \emph{energy density} and \emph{flux} they observe is important to consider the back-reaction in the semi-classical field equations. This can be motivated by a look at \fig{Kruskaldiag} where we have a gravitationally collapsing system and hence the relevant part of the Kruskal extension. Now consider two rays $\mathit{1}$ and $\mathit{2}$ as straddling the horizon at $r=2M$, inside and outside respectively. The outside ray reaches the asymptotic infinity and the redshift gives rise to the standard Hawking effect while the inside ray goes and hits the singularity in the finite time. It is not clear immediately what happens for the rays hitting the singularity and hence one can ask whether there is any accumulation of flux near the singularity. These questions give a sufficient purpose to look at the insides. We would like to note that although the result concerning the flux at the horizon crossing has been reported earlier \cite{ms2013}, it is important to look at its evolution during the complete history of the in-fall - outside, through and inside the horizon and also how it compares with the flux observed by the static observers at various radii along the in-fall. 

We take up these issues in this paper working in the framework of reference \cite{ms2013}. The key feature therein is the use of the new globally defined null coordinates. These coordinates are regular everywhere as well as on the horizon except at the singularity as we will see briefly in \sec{sec:coordinatesetup}. There is a natural and uniquely defined ~~``in"- vacuum for the scalar field in this background\footnote{It is to be noted that this is in contrast to working in the Kruskal extension of Schwarzschild metric where different vacua need to be defined according to the different boundary conditions (See for example \cite{Kim01,Kim02,Kim03}). We believe that the in-vacum is more natural to work with in the collapse geometry.}. We work out the renormalised stress-energy tensor for this scalar field in \sec{sec:tab} and look at the observables - energy density and flux - measured by the static observers and radially in-falling observers for different cases in \sec{sec:observables} before concluding in~\sec{sec:conclusion}.  

As a side note we would like to mention the ongoing debate concerning the firewalls at the horizon. AMPS~\cite{Almheiri13} have suggested the presence of a ``firewall" at the horizon according to principle of information conservation and various quantum constraints. But some others (see \cite{Unruh76,Wilburn11} for example) claim that the in-falling observer should not detect anything near the horizon on the basis that the Hawking effect reduces to the Unruh effect in the near horizon limit. It would help to know what is happening for the in-falling observers near the horizon at least semi-classically. 

\begin{figure}[b!]
\centering
\includegraphics[scale=1]{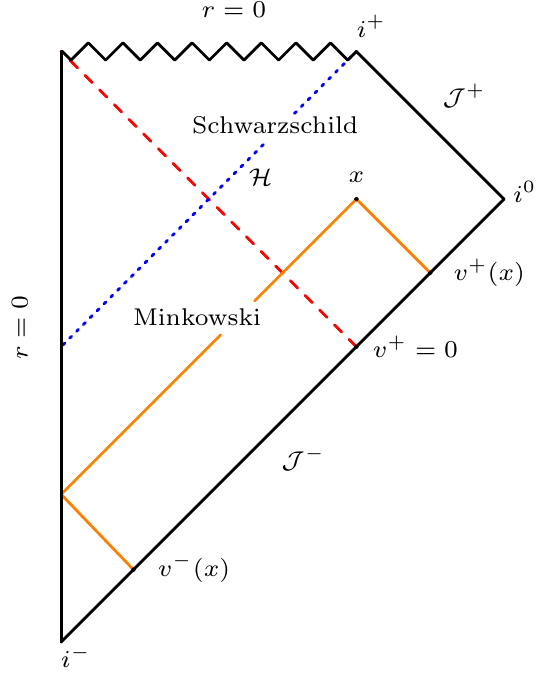}
\caption{(Color online) Penrose diagram for the Vaidya spacetime with two regions - Minkowski (interior) and Schwarzschild (exterior) separated by null thin-shell (dashed) at $v = 0$. The dotted line denotes the event horizon $\mathcal{H}$. Any event $x$ on the spacetime can be labelled by two globally defined null coordinates $(v^+,v^-)$ which correspond to a null ray coming directly from $\cal{J}^-$ and a null ray traced back in the past to the vertical line ($r=0$) and then reflected off to $\cal{J}^-$ respectively.}
\label{penrosediag}
\end{figure}

\section{The Gravitational Collapse Geometry}
\label{sec:coordinatesetup}

For simplicity, we shall work with a collapsing, thin-null shell since the results in any realistic collapse are similar as far as the structure is concerned. We thus have a collapsing null thin-shell separating the spacetime into two regions:  the Minkowski interior and a Schwarzschild exterior. The metric for this case can be expressed in Eddington-Finkelstein advanced coordinates $(v,r)$ as
\be\label{vaidya}
ds^{2}=-\left(1-\f{r_{s}}{r}\Theta(v)\right)dv^{2}+2dv dr+r^{2}d\Omega^{2}
\ee
where $r_{s}=2M$ defines the event horizon with the surface gravity $\kappa = 1/2r_{s}$, the Heaviside function $\Theta(v)$ separates the two regions - Minkowski and Schwarzschild mentioned above and $d\Omega^{2}=d\theta^{2}+\sin^{2}\theta d\varphi^{2}$ is the standard angular part. It is however convenient to introduce globally defined null coordinates $(v^+,v^-$) for the $(v,r)$ sector of the spacetime as follows: For each event $x$ (see \fig{penrosediag}), we can have two null rays one incoming from $\cal{J}^-$ which labels it $v^+$ and another is the outgoing which we can track back to $\cal{J}^-$  after reflecting off the origin in the past $r= 0$, which is the vertical line, labelling it $v^-$ which naturally makes $v^+\geq v^-$. This coordinate construction is regular everywhere except at the singularity and is generic irrespective of the nature of collapse. The mapping $(v,r)\mapsto(v^{+},v^{-})$ for the Vaidya metric (\eq{vaidya}) is straightforward (the details can be found in \cite{ms2013}) giving $v^+ = v$ and  
\begin{multline}\label{va}
v^{-}(v,r)=
\left
\{
\begin{array}{c}
v-2r\quad\quad\,\quad \qquad \quad \quad\quad\textrm{for}\ v<0\\
-2r_{s}\Big[1+W\left(\delta\,e^{\delta-\kappa v}\right)\Big]\quad\textrm{for}\ v\geq0\end{array}
\right.
\end{multline}
where $\delta\equiv r/r_{s}-1$ and $W(z)$ is the Lambert $W$-function. Since $W(z)\sim z$ as $z\rightarrow0$, we have the equation of the event horizon as $v_{-}=-2r_{s}$. With the above mapping of coordinates we can express the metric in \eq{vaidya} in a more pertinent form:

\be
\label{nullmetric}
ds^2 = - C(v^+,v^-) dv^+dv^- + r^2d\Omega^2
\ee
where $C(v^+,v^-)$ is the conformal factor of the (1+1) sector given by
\be
\label{conformalfactor}
C(v^+,v^-) =  1+\left(-1 + \f{a-1}{a}\f{W(-a e^{-a + \kappa v^+})}{1+W(-a e^{-a + \kappa v^+})}\right)\Theta(v^+)
\ee
with $a \equiv 1 + \kappa\,v^-$. 

\section{Introducing the scalar field}
\label{sec:scalarfield}

We consider minimally coupled massless scalar field on the background in \eq{nullmetric}. It is well known that the dominant contribution to the Hawking effect comes from the $s$-waves ($l=0$ part of the spherical wave expansion). This reduces the dynamics to effectively (1+1) dimensions and the solution to the field equation can be written down as, 
\begin{equation}
\frac{\phi (x)}{r}\sim \lim _{r\rightarrow \infty}\frac{\phi (v^{+},r)-\phi (v^{-},r)}{r}
\end{equation}  
The physical interpretation of the $(v^{+},v^{-})$ coordinates is also tied to this eikonal 
approximation. The field at a given point $x$ is given as a superposition of two spherical waves originating from past null infinity: a convergent wave arriving directly from $v^{+}$ and a divergent wave arriving from $v^{-}$ after a reflection off the origin. The vacuum for this field is defined naturally and uniquely at $\mathcal{J}^-$ and hence the name ``in"-vacuum. Reduced to two dimensions, the dynamics gets coded in the conformal factor and we shall now move on to the calculation of renormalized stress-energy tensor for this scalar field.

\section{Renormalized stress-energy tensor}
\label{sec:tab}

We follow the standard procedure (see \cite{davies1976, brout1995} for details) for evaluating the stress-energy tensor for a scalar field in the conformal metric which gives the components in terms of the conformal factor as,
\begin{align}
\label{EMT03}
\langle \hat{T}_{++}\rangle &= -\f{1}{12\pi}C^{1/2}\partial ^{2} _{+}C^{-1/2}\nn\\
\langle \hat{T}_{--}\rangle &= -\f{1}{12\pi}C^{1/2}\partial ^{2} _{-}C^{-1/2};\nn\\
\langle \hat{T}_{+-}\rangle &= \f{1}{96\pi}\partial _{+}\partial _{-}\ln C
\end{align}
where we have employed the notation $\pm$ as shorthand for coordinates $v^{\pm}$ respectively. Given the conformal factor in \eq{conformalfactor}, we have the following list of derivatives, 
\begin{align}
\partial _{+}C=&\left(-1+\frac{a-1}{a}\frac{W}{1+W}\right)\delta (v^+)+\left(\frac{a-1}{a}\right)\nn\\
&\frac{\kappa  W}{(1+W)^{3}}\Theta (v^+)\nn
\end{align}
\begin{align}
\partial ^{2}_{+}C=&\left(-1+\frac{a-1}{a}\frac{W}{1+W}\right)\delta '(v^+)+2\kappa\left(\frac{a-1}{a}\right)\nn\\
&\hspace{-1.0cm}\frac{W}{(1+W)^{3}}\delta (v^+) +\left(\frac{a-1}{a}\right)\frac{\kappa^2 W(1-2W)}{(1+W)^{5}}\Theta (v^+)\nn\\
\nn\\
\partial _{-}C = & \frac{1}{a^{2}}\frac{W}{1+W}\left(1-\left(\frac{a-1}{1+W}\right)^2\right)\kappa\Theta (v^+)\nn\\
\nn\\
\partial ^{2} _{-}C = &\left(-\frac{2}{a^{3}}\frac{W}{1+W}-\frac{a-1}{a^{3}}\frac{W}{(1+W)^{3}}+\left(\frac{a-1}{a}\right)^{3}\right.\nn\\
&\left. \hspace{-0.5cm}\frac{W(1-2W)}{(1+W)^{5}}-\frac{(a-1)}{a^{3}}\frac{2W}{(1+W)^{3}}\right)\kappa ^{2}\Theta (v^+)\nn \\
\nn\\
\partial^2_{+-}C = &\left(\frac{1}{a^{2}}\frac{W}{1+W}-\left(\frac{a-1}{a}\right)^{2}\frac{W}{(1+W)^{3}}\right)\kappa \delta (v^+)
\nn\\
&\hspace{-1.0cm}-\left(\frac{1}{a^{2}}\frac{W}{(1+W)^{3}}-\left(\frac{a-1}{a}\right)^{2}\frac{W(1-2W)}{(1+W)^{5}}\right)\kappa ^{2} \Theta (v^+)\nn
\end{align}
where again note that $a=1+\kappa v^-$ and $W = W(-a e^{-a+\kappa v^+})$. Though these expressions seem complicated in the complete forms, they are simple if confined to individual regions. We observe that for $v^+<0$ all the components of stress-energy tensor vanish as they should being in the Minkowski region. Then there is a discontinuity at $v^+=0$ due to the discontinuity in the metric itself and we have factors of $\delta(v^+)$ which contribute only at $v^+= 0$. For $v^+>0$, the components of the stress-energy tensor are simply,
\begin{align}
\label{tabcomponents}
\langle \hat{T}_{--}\rangle &=\kappa ^{2}\frac{\left(1+4W\right)\left(1+\delta \right)^{4}-\left(1+W\right)^{4}\left(1+4\delta\right)}
{48\pi \left(1+W\right)^{2}W^{2}\left(1+\delta \right)^{4}}\nn\\
\langle \hat{T}_{++}\rangle&=-\frac{\kappa ^{2}}{48\pi}\frac{1+4\delta}{\left(1+\delta \right)^{4}}\nn\\
\nn\\
\langle \hat{T}_{+-}\rangle&=\frac{\kappa ^{2}}{12\pi}\frac{1+W}{W}\frac{\delta}{\left(1+\delta\right)^{4}}
\end{align}
in terms of $\delta$ using \eq{va} with $W=W(\delta e^{\delta -\kappa v^+})$. This is for future convenience since we need to look at observables at different radii concerning relevant observers. 

\begin{figure}[t!]
\includegraphics[scale=1]{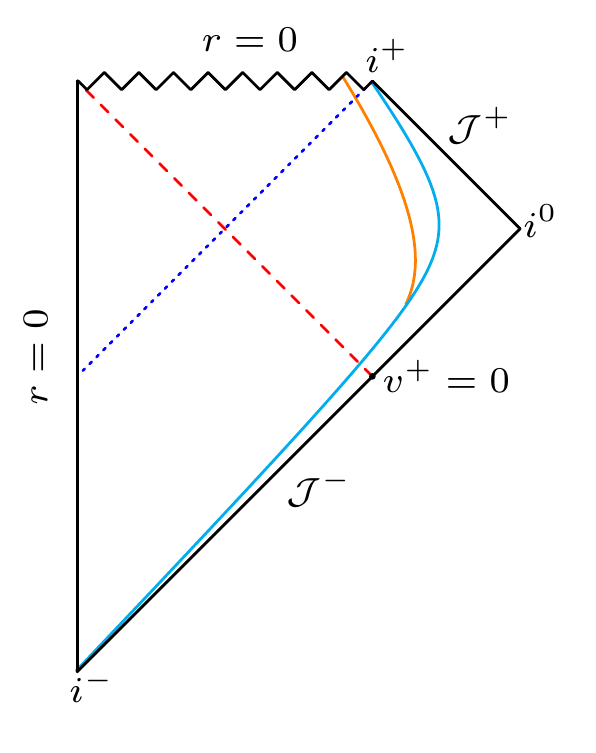}
\caption{(Color Online.) The trajectories of Static observer (cyan) and a radially in-falling observer who diverts from the static path at some time after $v^+>0$ (orange).}
\label{penrosediagtraj}
\end{figure}


\section{Energy density and Fluxes}
\label{sec:observables}

Knowing the stress-energy tensor, we can construct the local observables, energy density and flux, following reference \cite{ford1993,Paranjape2013}. Let $u^a$ be the $4$-velocity (or $2$-velocity in our case) of an observer and $n^b$ be a normal such that $u_a n^a = 0$. Then we have,
\begin{align}
U &\equiv \langle \hat{T}_{ab} \rangle u^{a}u^{b}\nn\\
F &\equiv - \langle \hat{T}_{ab} \rangle u^{a}n^{b}
\label{observables}
\end{align}
as the local energy density and flux perceived (in the direction given by $n^a$) respectively by that particular observer at any point along its world line. We shall now consider these quantities for the static observers and free radially in-falling observers for various trajectories of the type as shown in \fig{penrosediagtraj}.  
\begin{figure*}[t!]
\includegraphics[width=0.33\textwidth]{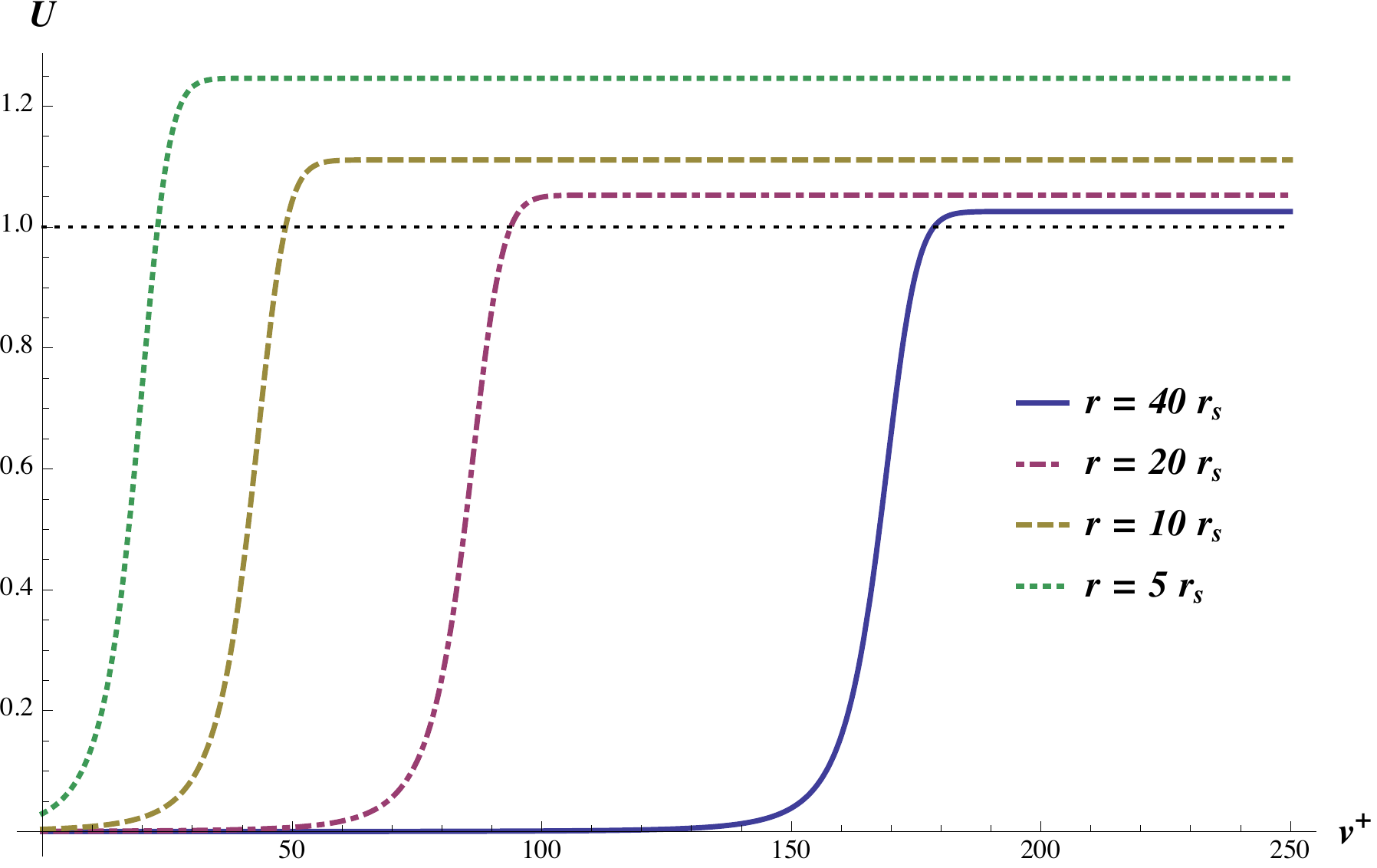}\hfill
\includegraphics[width=0.33\textwidth]{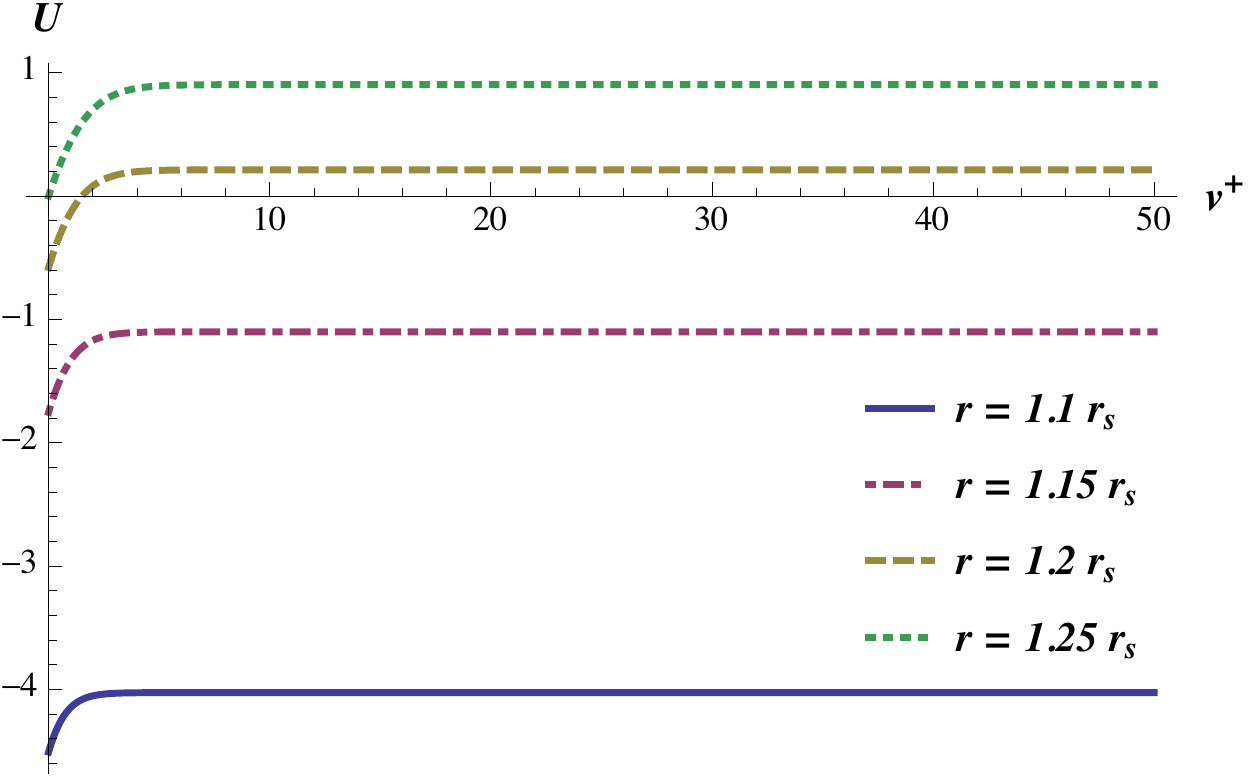}\hfill
\includegraphics[width=0.33\textwidth]{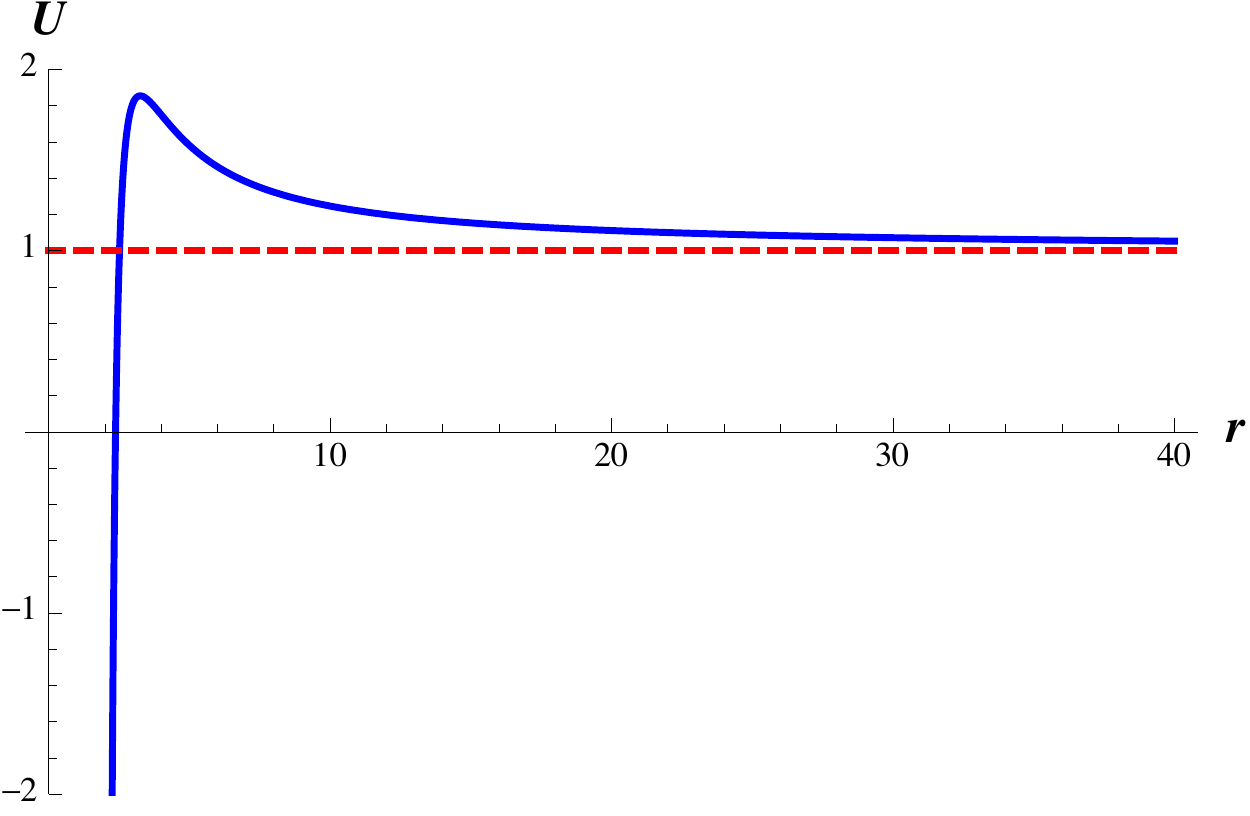}
\includegraphics[width=0.33\textwidth]{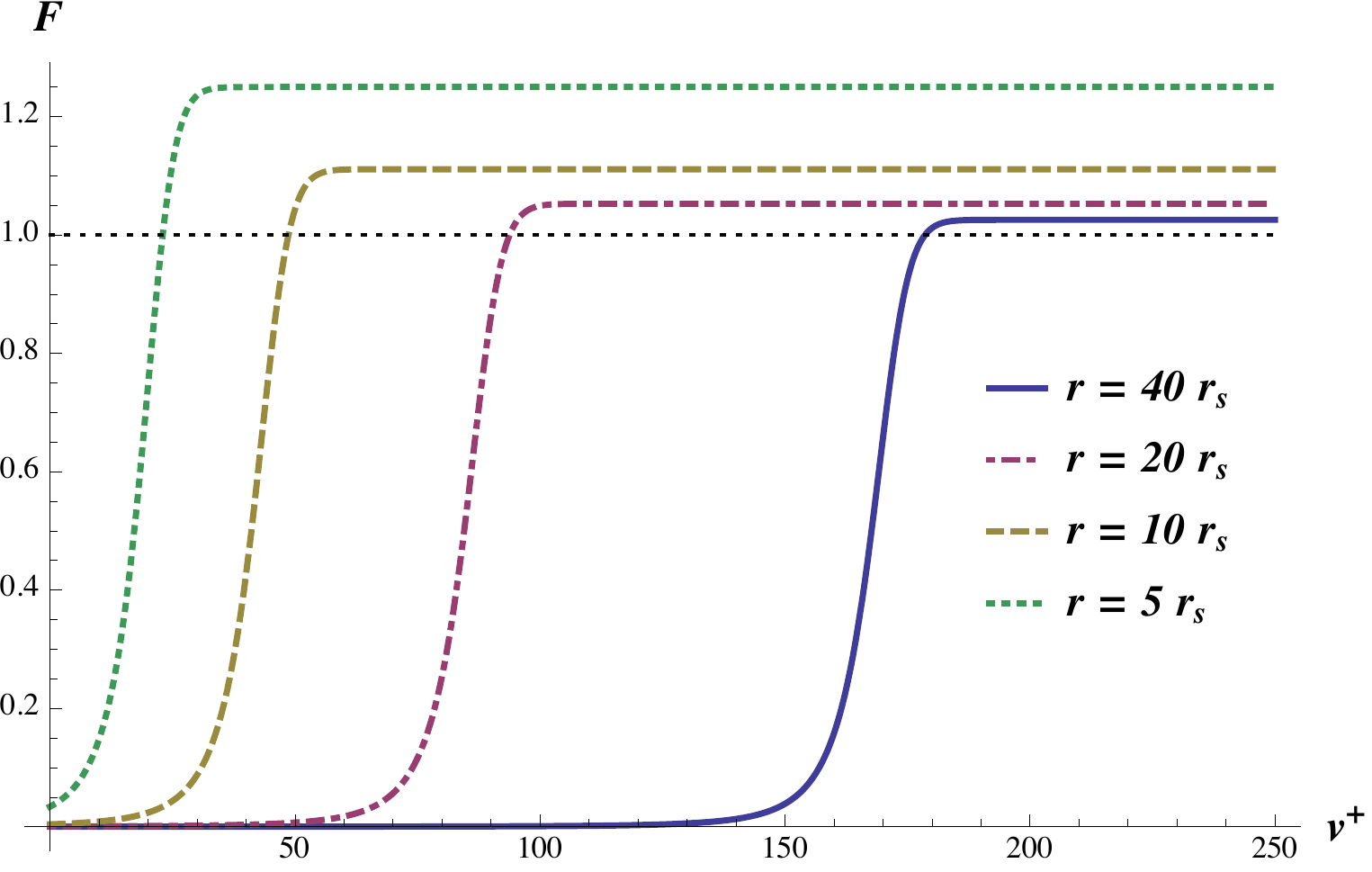}\hfill
\includegraphics[width=0.33\textwidth]{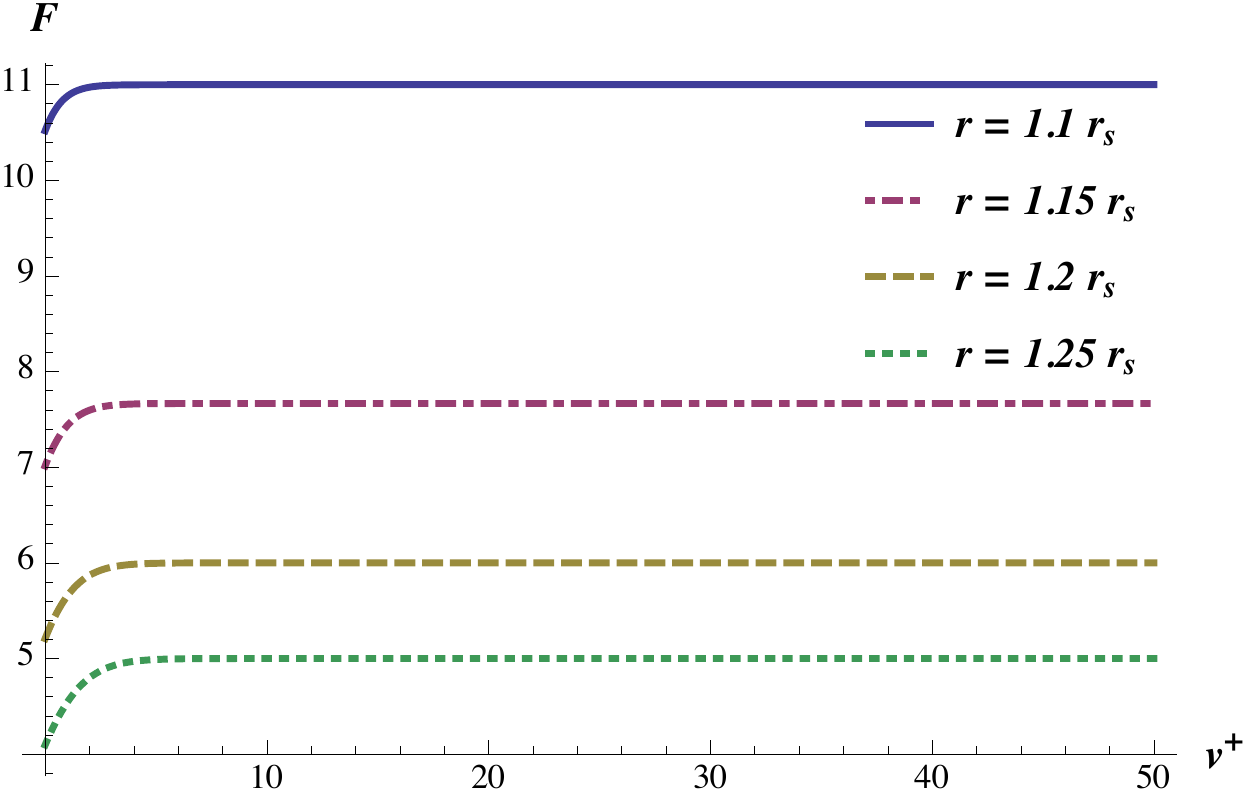}\hfill
\includegraphics[width=0.33\textwidth]{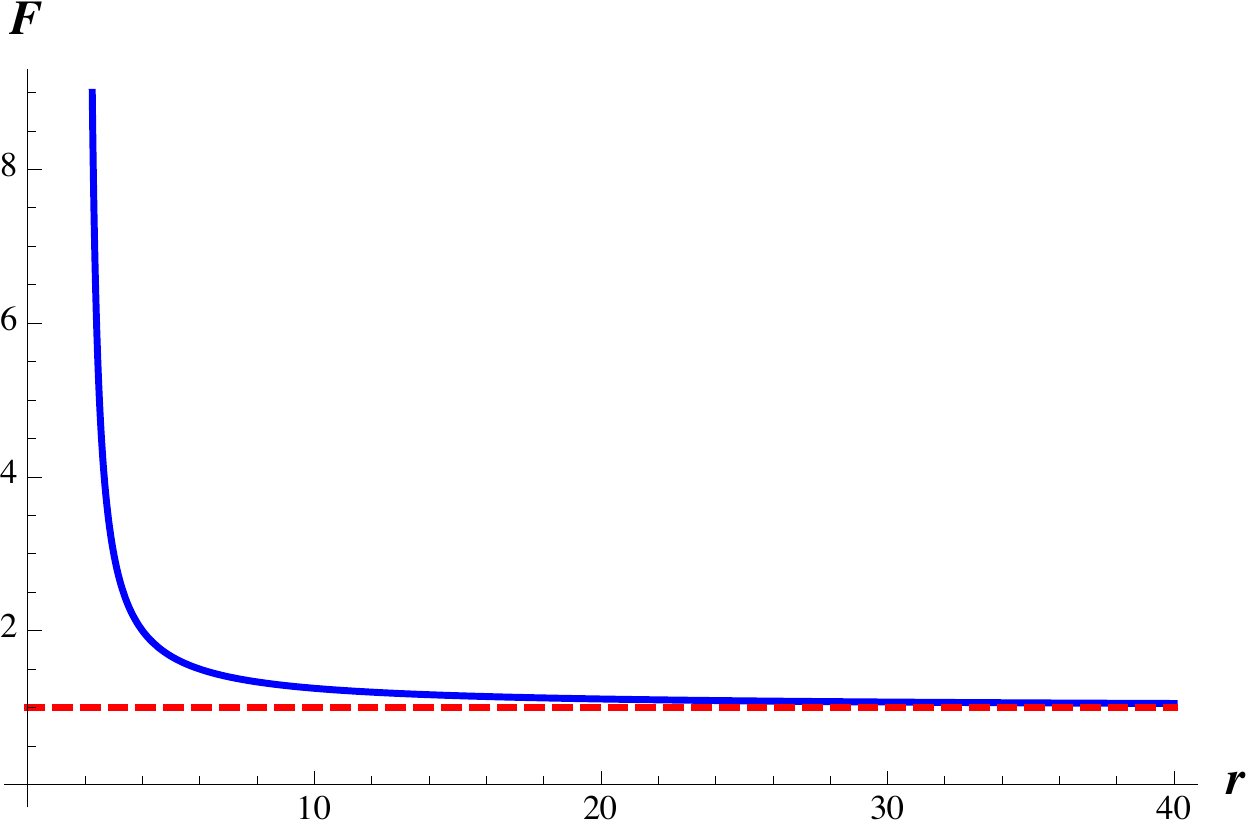}
\caption{(Color online) The figures show energy densities, $U$ and fluxes, $F$ perceived by a static observers specified by their radial positions along their trajectory as a function of $v^+$ which is proportional to the proper time. Both $U$ and $F$ are normalized to their value at infinity, $\pi T_H^2/12$ and we have taken $M=1$ so that $r_s = 2$ in our case. The energy density and flux show similar behaviour of growing slowly early on and later making a transition to saturated thermal state for larger radii but as we near $r_s$ the differences come in. The energy density can be negative for region below certain radius (see text for details). Then we have the late-time energy density and flux for static observers as a function of $r$ showing that the energy density increases with decreasing $r$ to a maxima, then decreases thereon becoming negative at a point and diverging at $r = r_s$. The flux on the other hand is positive throughout increasing with decreasing $r$ and diverging at $r = r_s$.}
\label{UFstatic}
\end{figure*}

\subsection{Static Observers}

Consider an observer at rest in the Minkowski region specified by a constant radial position, $r > r_s$. The observer can stay put at that radius till the shell hits it at $v^+ = 0$. After this point, the observer has to fire rockets or accelerate in order to be stationary at the that fixed radius. The velocity components are then,
\begin{align}\label{dvs}
\dot{v}^+&= \sqrt{\frac{1+\delta}{\delta}}\nn\\
\dot{v}^{-}&= \frac{W(\delta e^{\delta -\kappa v^+})}{1+W(\delta e^{\delta -\kappa v^+})}\sqrt{\frac{1+\delta}{\delta}}
\end{align}
for a given $\delta$ and the components of the normal with $n_a\dot{v}^a = 0$ are 
\begin{align}\label{stat07}
n^{+} = \dot{v}^+ ;\,\,\,  n^{-} = -\dot{v}^- 
\end{align}
which can easily be verified to be the outward radial normal. Then, for $v^+ > 0$, the energy density and flux can be computed using \eq{observables} for static trajectories as a function of $v^+$ which is proportional to the proper time along them and are plotted in \fig{UFstatic}. (An analysis on similar lines concerning radiation from collapsing shells for observers outside the horizon was also done in \cite{Paranjape2013}). The energy density and flux show similar behaviour for static trajectories at large radii. The onset begins from positive values and grows in a very slow linear way and then makes a sharp transition and saturates thereon. This can be tracked to the behaviour of Lambert W-function, so that the sharp transition occurs when the argument of $W(\delta e^{\delta -\kappa v^+})$ is close to unity, that is, when $v^+ \sim (\delta + \log\delta)/\kappa$, for $\delta>1$. For $\delta<1$, the transition time is given by $v^+ \sim \delta$. This shows that the thermalizing time is very small near the horizon where $\delta\rightarrow0$. The analytical expressions for the saturating value can be obtained in the late-time ($v^+$) limit when $W(\delta e^{\delta -\kappa v^+})\sim \delta\, e^{\delta -\kappa v^+}$ giving      
\begin{subequations}
\begin{align}
U &= \f{\pi T_H^2}{12}\left(1-\f{2\, r_s^4}{r^4}\right)\left(1-\f{r_s}{r}\right)^{-1} \\
F &= \f{\pi T_H^2}{12}\left(1-\f{r_s}{r}\right)^{-1}.
\end{align}
\end{subequations}
These expressions reduce to the Hawking energy density and flux for the asymptotic observers with $r \rightarrow \infty$ and otherwise the temperature is given by the Tolman blueshift factor. This factor diverges as one nears $r=r_s$.  As seen in \fig{UFstatic}, the energy density $U$ (at late-times) as a function of $r$ shows a growth to the maximum positive value as $r$ decreases to $1.63\,r_s \approx 3.25 M$ after which it decreases with $r$, vanishing at $r \approx 1.18\, r_s = 2.38 M$ and then becomes negative, subsequently diverging at $r= r_s = 2M$ implying that that there is a negative energy density region outside and near the horizon for static observers. The flux on the other hand is positive throughout and diverges at the horizon. Since the energy density is negative in the near- horizon region, it is better to interpret the flux in the same region as ingoing negative energy flux than a positive outgoing one. We shall now consider the case of radially in-falling (geodesic) observers.     

\subsection{Radially In-falling Observer}
 An in-falling observer in Schwarzschild metric is characterized by their energy, $E$ or initial radius, $r_i$ from which the free-fall begins. These two are related by
\be
\label{energy}
E = \left(1 - \f{r_s}{r_i}\right)^{1/2}.
\ee
Using the geodesic equation, we find the components of the velocity for the ingoing observer as
\be
\label{dv}
\dot{v}^+=\frac{E-\sqrt{E^{2}-\delta /1+\delta}}{\delta/1+\delta}\\
\ee
\be
\label{dvm}
\dot{v}^-= \frac{W(\delta e^{\delta -\kappa v^+})}{1+W(\delta e^{\delta -\kappa v^+})}\f{E+\sqrt{E^{2}-\delta/1+\delta}}{\delta/1+\delta}
\ee
and the components of the normal given by $n_a\dot{v}^a = 0$ are 
\begin{align}
n^{+} = \dot{v}^+;\,\,\,n^{-}=-\dot{v}^-
\end{align}
\begin{figure*}[t!]
\includegraphics[width=0.33\textwidth]{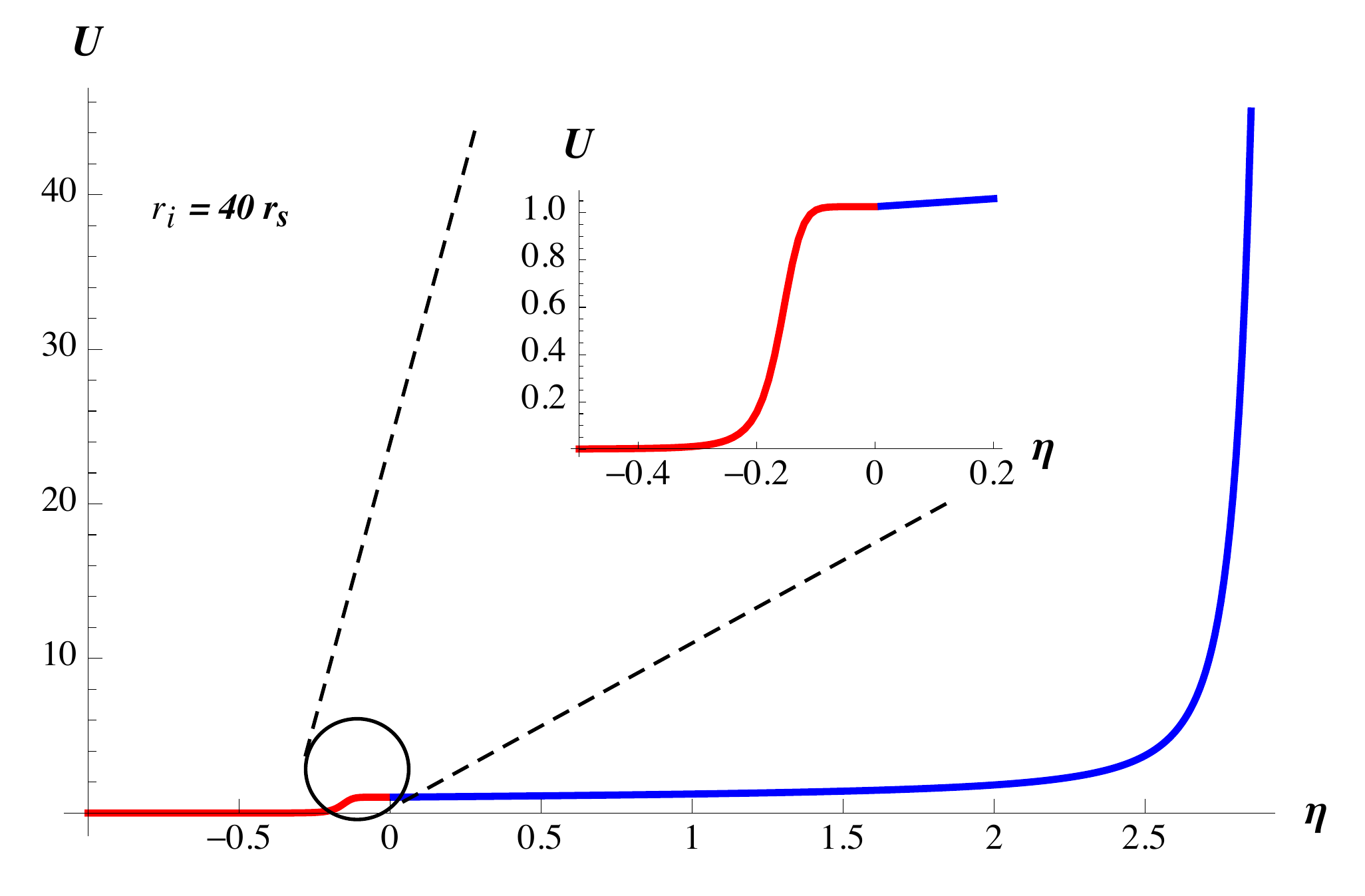}\hfill
\includegraphics[width=0.33\textwidth]{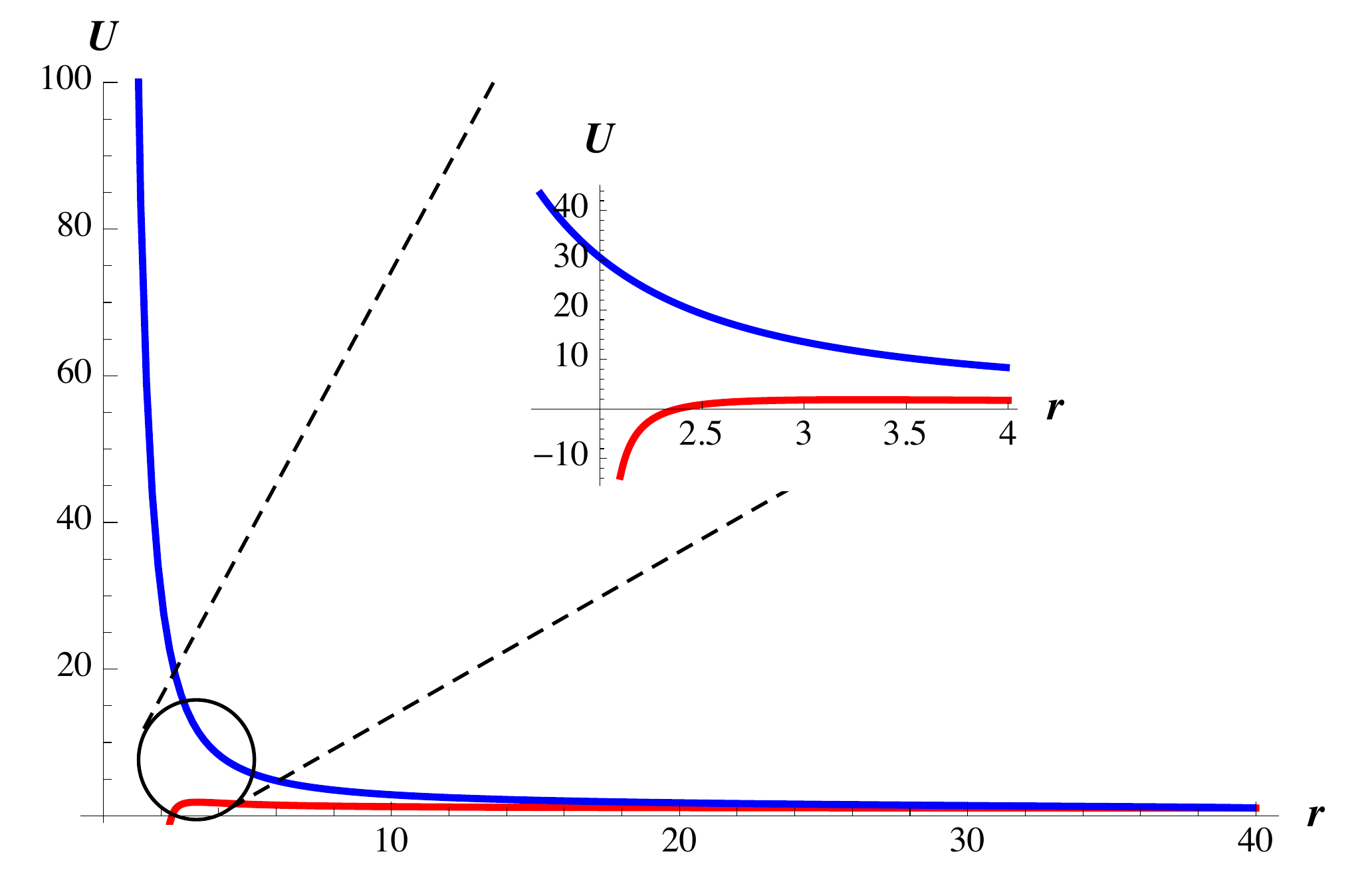}\hfill
\includegraphics[width=0.33\textwidth]{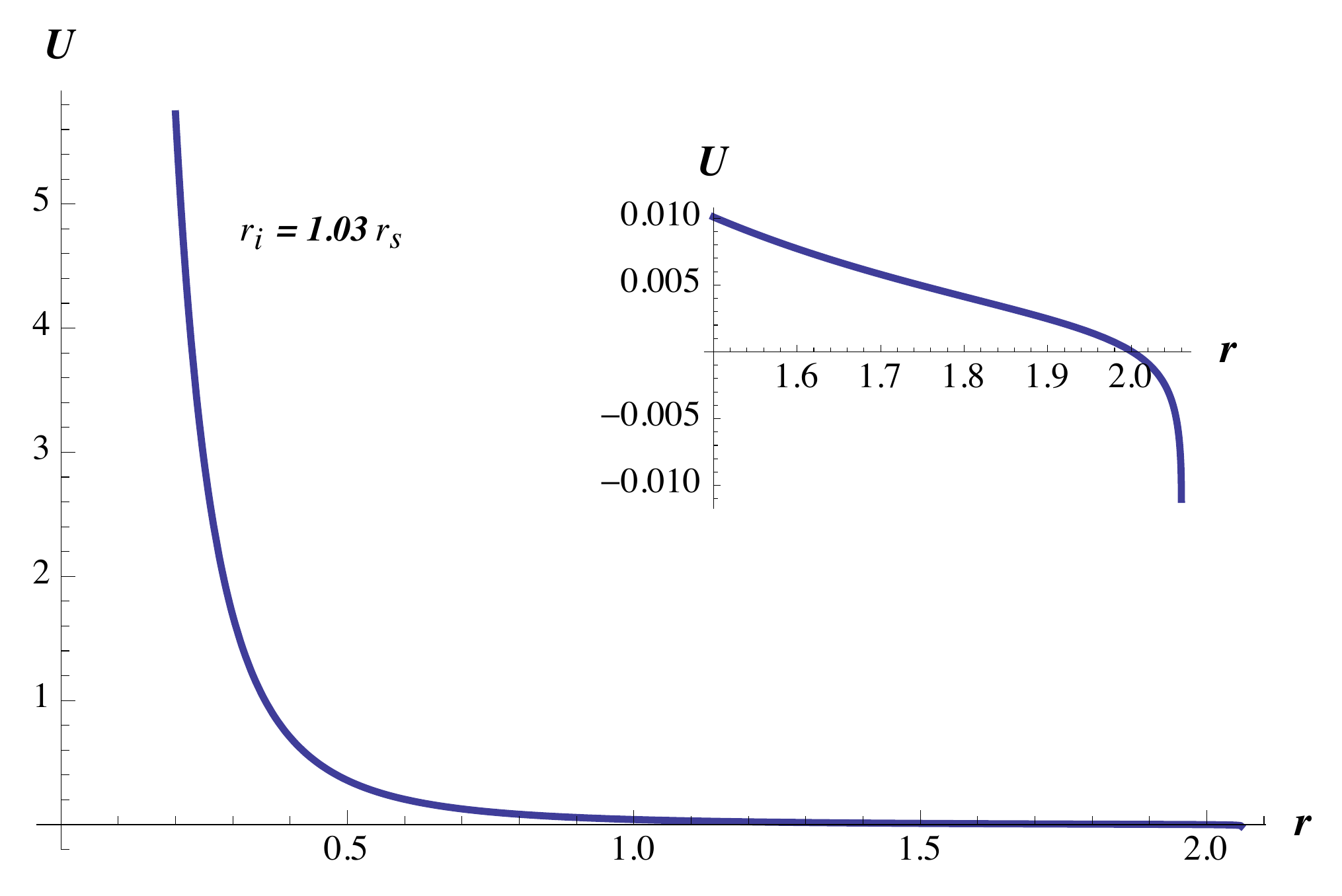}
\includegraphics[width=0.33\textwidth]{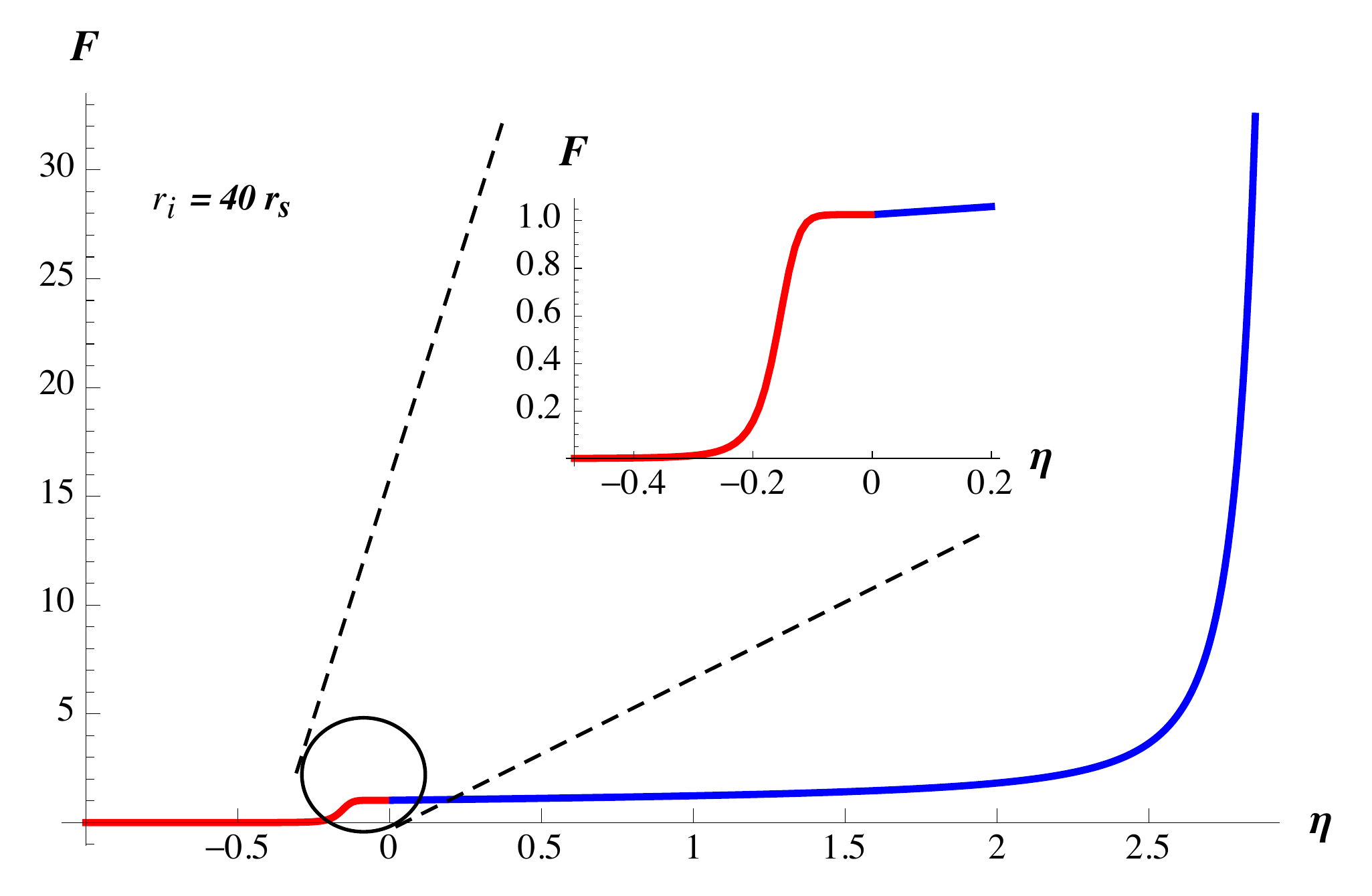}\hfill
\includegraphics[width=0.33\textwidth]{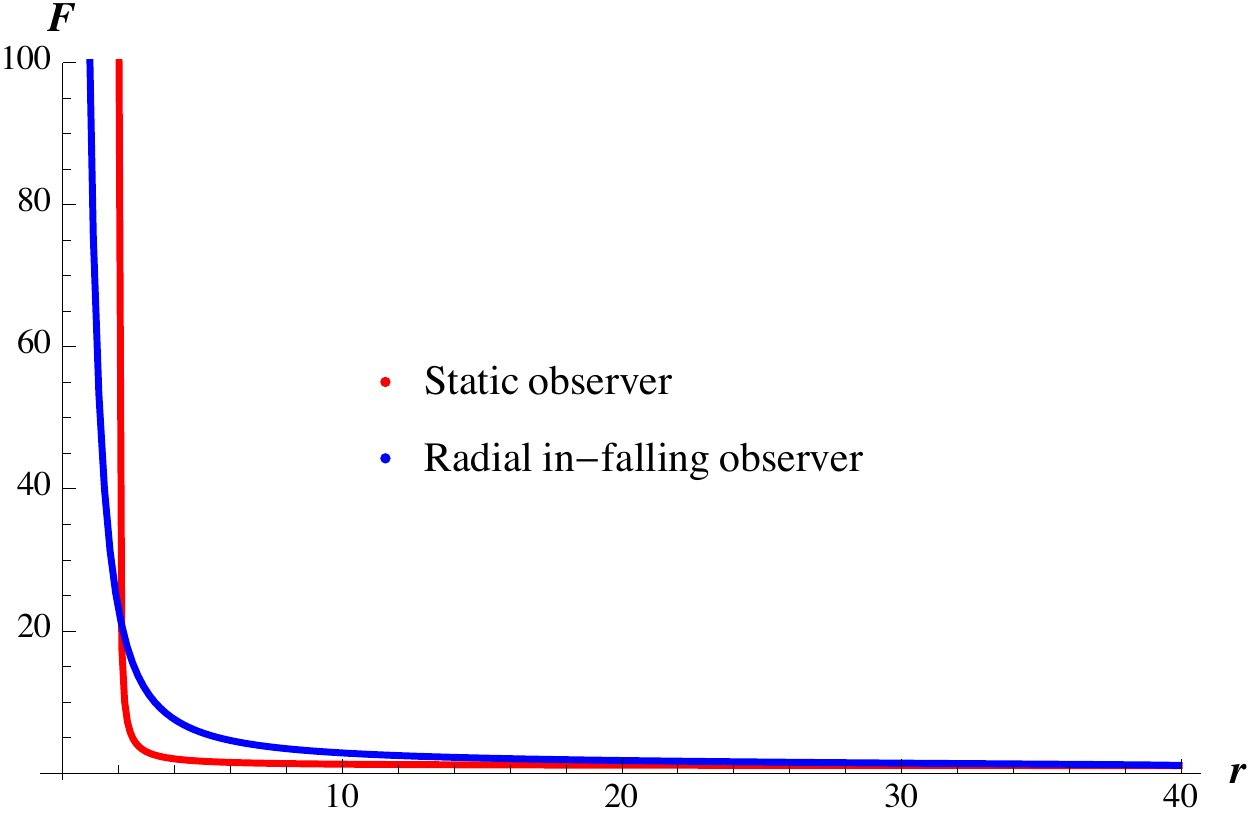}\hfill
\includegraphics[width=0.33\textwidth]{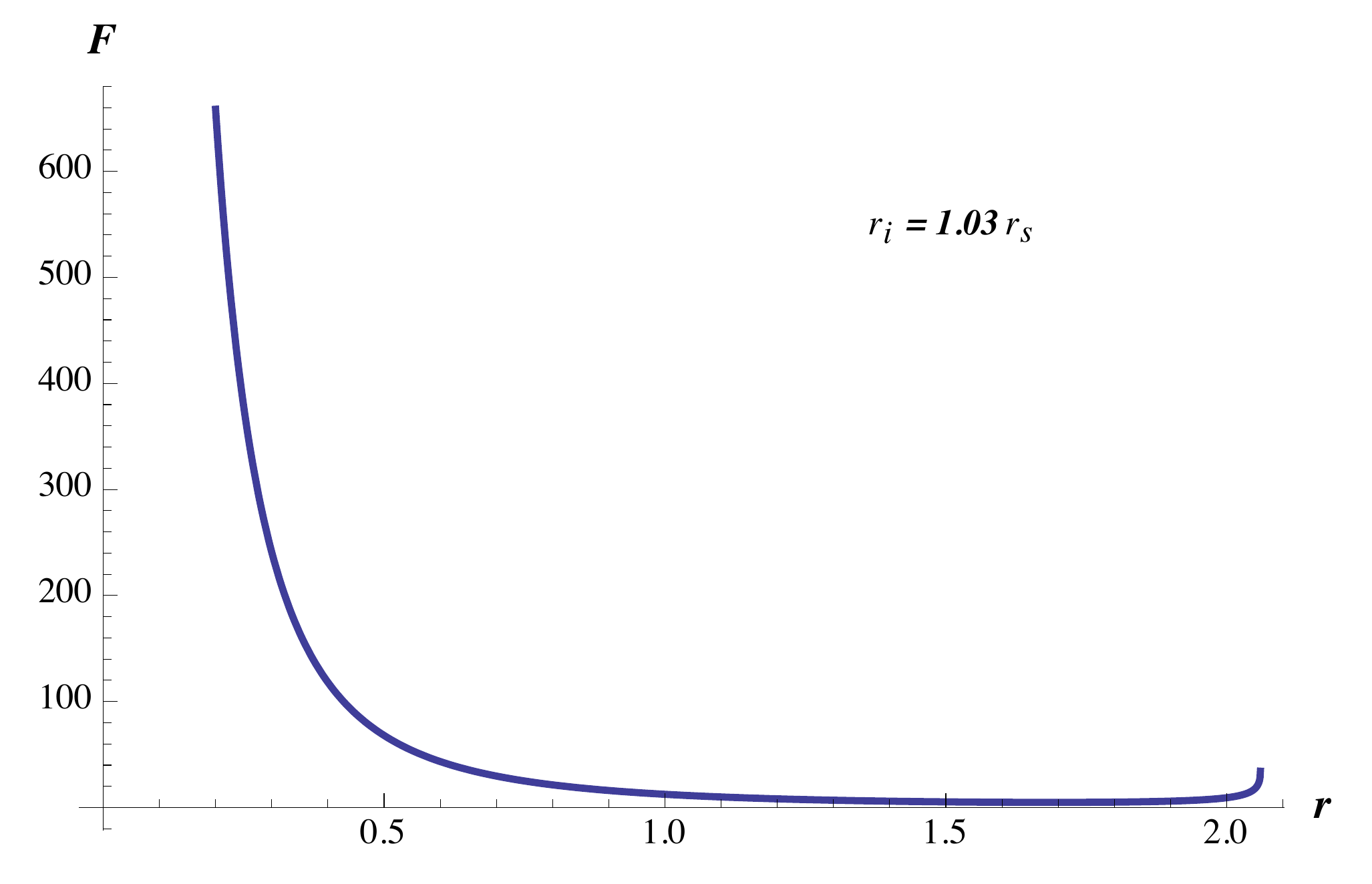}
\caption{(Colour online) The plots show Energy density (U) and Flux (F) perceived by a radially in-falling observer normalized by $\pi T_H^2/12$ and with $M=1$ so that $r_s = 2$. The first column gives both $U$ and $F$ as a function of parameter $\eta$ along the trajectory with the in-fall beginning at $r_i = 40\,r_s$. The second column compares $U$ and $F$ for the in-falling observer from $r_i = 20\,r_s$ with those observed by the thermal static observers at different radii. The last column shows $U$ and $F$ for in-fall beginning close to the horizon at $r_i = 1.03\,r_s$. (See text for description).}
\label{UFinfall}
\end{figure*}  
The sign of the normal is chosen such that the flux is positive at the start of infall matching with the observer who is on a static trajectory at $r=r_i$. We also need the evolution of $v^+$. To match with the static observer, we note by integrating the first equation of \eq{dvs} that, 
\be
v^+ (\eta) = \tau_i (1+\eta)\left(\f{1+\delta_i}{\delta_i}\right)^{1/2} = \f{\tau_i}{E}\,(1 + \eta) 
\ee  
where we have introduced a parameter $\eta$ such that $v^+(0) = v^+_i = \tau_i/E$. The parameter $\eta$ is proportional to proper time $\tau$ on the static trajectory. For the infall, the solution of geodesic equations is given by,
\begin{align}
\label{rad04}
r &= \f{r_i}{2}(1+\cos \eta)\nn\\ 
\tau &= \tau_i + \f{r_i}{2}\sqrt{\f{r_i}{r_s}}\left(\eta +\sin \eta \right)
\end{align}
with $0\leq\eta\leq\pi$ such that $r(\tau_i) = r (\eta = 0) = r_i$. The first equation in the above set can be inverted to get
\be
\label{eta}
\eta = \cos^{-1}\left(\f{2r}{r_i} - 1\right)
\ee
\\
which is useful whenever we need everything as a function of $r$. Using the second equation we have,
\be
\f{\mathrm{d}\tau}{\mathrm{d}\eta} = \f{r_i}{2}\sqrt{\f{r_i}{r_s}}\,(1+\cos\eta)
\ee
which with the first equations of \eq{rad04} and \eq{dv}, gives
\be
\label{dveta}
\frac{\dd v^+}{\dd\eta} = r_i\cos\eta/2\f{\sqrt{r_i/r_s}(1- r_s/r_i)^{1/2}\cos\eta/2 - \sin\eta/2}{1- r_s/r_i \,\rm{sec}^2\eta/2}
\ee
that can be integrated to get $v^+(\eta)$ for $\eta\geq0$. Thus we have,
\be
v^+ (\eta) = v^+_i  + f(\eta) - f(0)
\ee
where $f(\eta)$ symbolically represents the integral of \eq{dveta}. The energy density and flux are then given by using the definition in \eq{observables} as $U(\eta, r_i,\tau_i)$ and $F(\eta, r_i,\tau_i)$ or as function of $r$ by using \eq{eta} to replace $\eta$. We give the complete expressions for these quantities in \app{app:A}. The two plots in the first column in \fig{UFinfall} show $U$ and $F$ as functions of $\eta$ for $r_i = 40\,r_s$ with $M = 1$ so that $r_s = 2$ normalized by $\pi T_H^2/12$. For $\eta \leq 0$, both the observables match with those of the static observer at $r_i$ which are thermalized to the values shifted by the Tolman factor. From then on, both the energy density and flux show a steady growth but remain finite even at the horizon crossing when $\eta_H \approx 2.824$ and diverge as the singularity is approached. We next compare the energy density and flux measured at different radii by the static observers at those locations and the radially in-falling observer. This time the plots are for the in-fall that begins at $r_i = 20 r_s$. Observe that both energy density and the flux grow steadily and are positive for the in-falling observer and are larger than what the corresponding static observers at different radii perceive except near the horizon, where the static $U$ and $F$ diverge. For the in-falling observer both the observables are regular and finite during the \emph{transition} through the horizon and diverge subsequently at the eventual singularity. In the third column, 
\begin{figure*}[t!]
\includegraphics[width=0.5\textwidth]{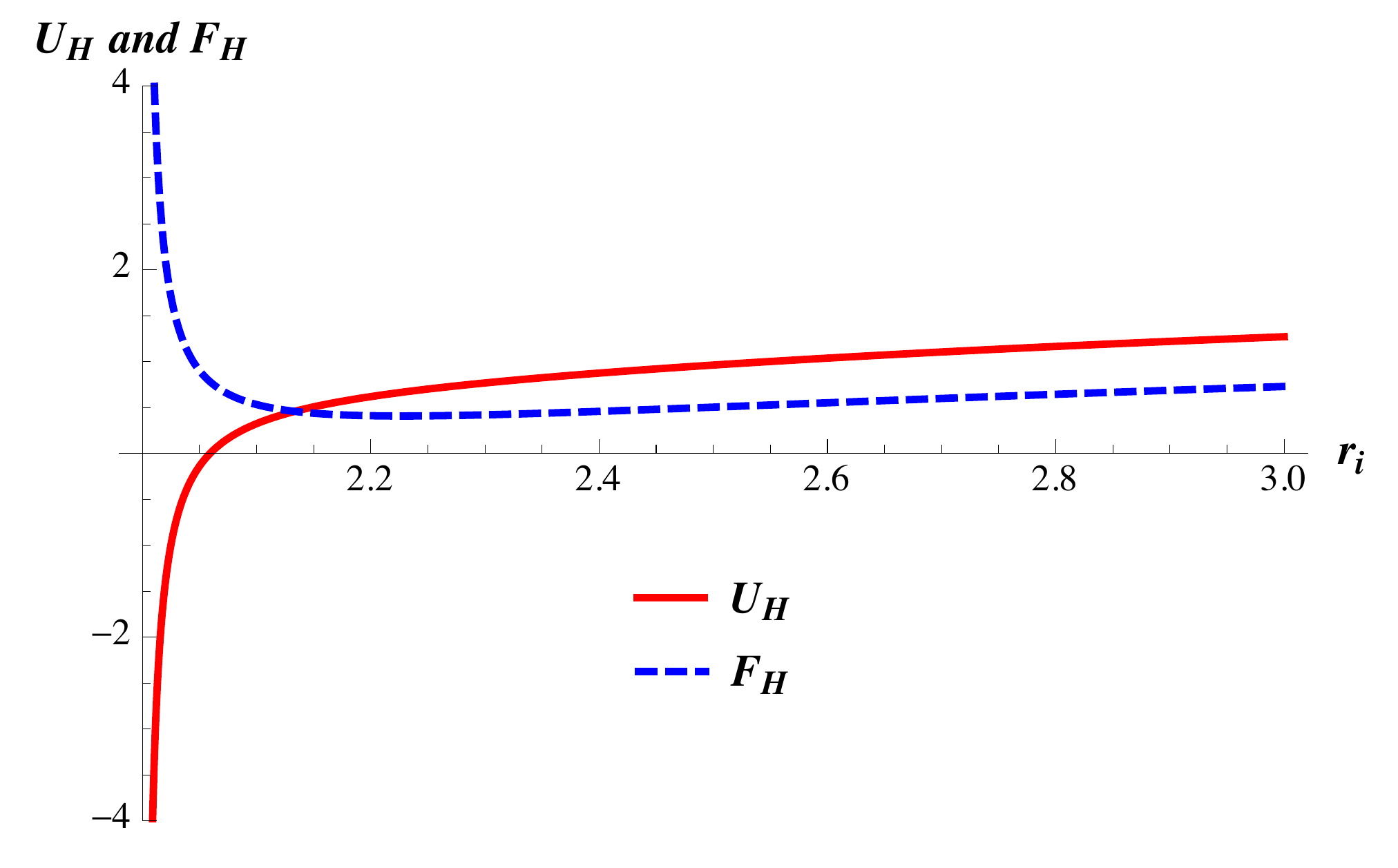}\hfill
\includegraphics[width=0.5\textwidth]{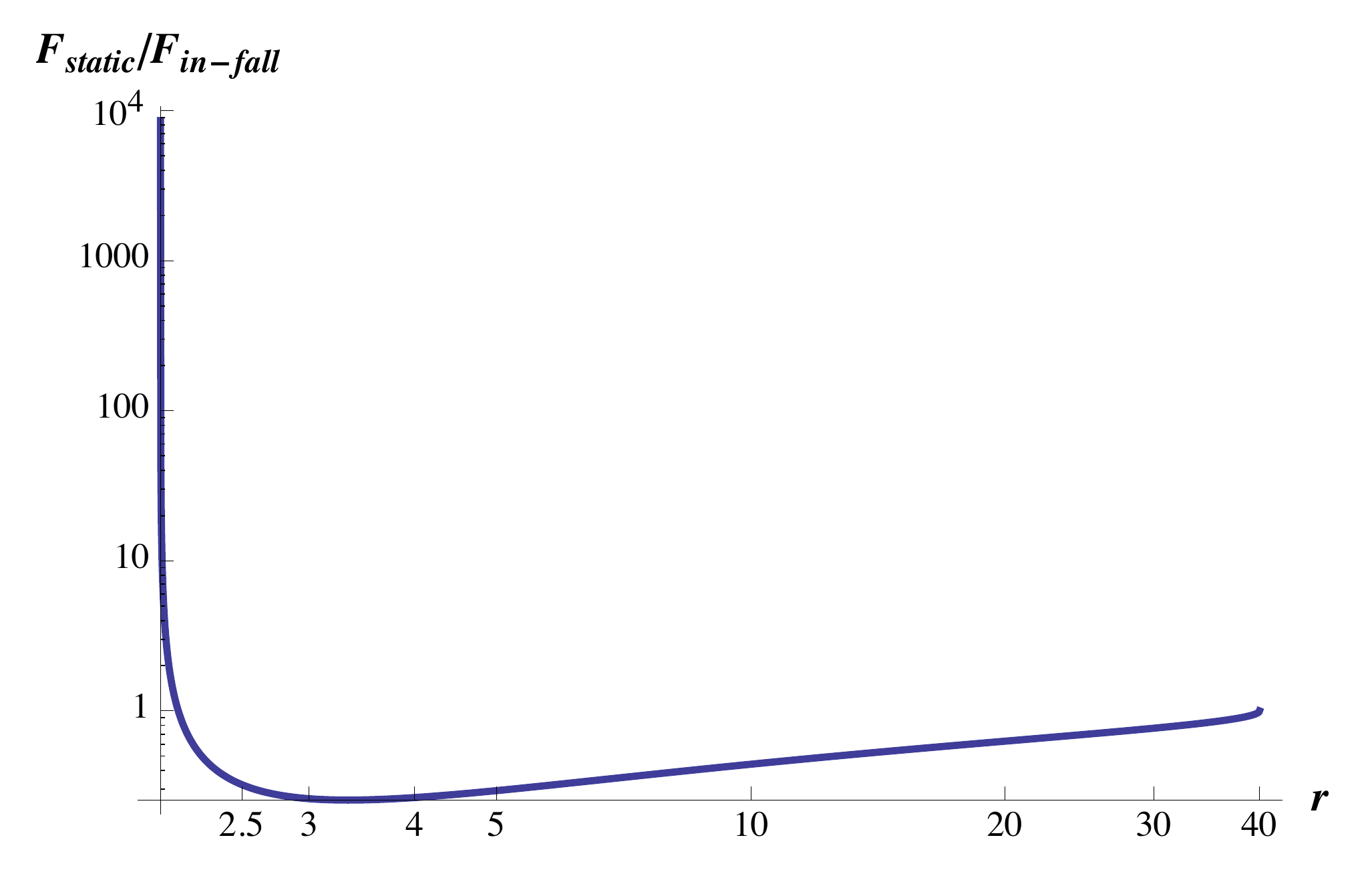}
\caption{(Colour online) The first plot shows the energy density (U) and flux (F) perceived by a radially in-falling observer at horizon crossing as a function of the initial radius of the in-fall. Next, we plot the ratio of the flux seen by a static observer at different radii to the flux seen by a radial in-faller.}
\label{UFHwithri}
\end{figure*}
we have $U$ and $F$ for an in-fall that begins close to the horizon for $r_i = 1.03 r_s$. In this case, we see that the energy density is negative (matching with the static case) early on, vanishes at the horizon and becomes positive after that, diverging at the singularity subsequently. The flux is positive early on and unusually shows a dip and then the usual growth pattern. The analytical expressions for the energy density and flux at horizon crossing can be obtained easily as  
\begin{align}
U_H &= \pi T_H^2\left(\f{2}{3} - \f{1}{48 E^2} + 2E^2\right)\nn \\
F_H &= \pi T_H^2\left(\f{1}{48 E^2} + 2E^2\right)
\end{align} 
in the late-time limit so that the factors of $\exp(-2\kappa v^+)$ drop away. For $E = 1$,  $U_H\approx 32 U_\infty$ and $F_H \approx 24\,F_\infty$. As is seen in reference \cite{ms2013}, the detectors on such a trajectory measure an ``effective" temperature of $T \approx 4T_H$ near the horizon (in the UV limit) but in a strict non-adiabatic regime. Thus, the temperature, \emph{if at all} inferred from this flux when normalized to the flux at infinity, does not match the one with the detector calculation.  We compare the energy density and flux at the horizon crossing plotting these expressions in \fig{UFHwithri} with the initial radius $r_i$ which is connected to energy $E$ by \eq{energy} characterizing the trajectory. Both the observables are regular except for the in-falls that begin very close to the horizon. The energy density at the horizon crossing \emph{vanishes} for the in-fall beginning at $r_i = 1.03\,r_s$, goes negative for the any in-fall that begins with $r_i$ below that and diverges for the in-fall beginning at the horizon. The flux is positive for all $r_i$ and also diverges for the in-fall beginning at the horizon. This is not too unsual since even for the static observer we have seen the same effect and any in-falling observer would thermalize quickly near the horizon to the value of the static observer. The regularity of the flux while near the horizon for the in-falling observer with respect to the static observer is also shown in the second plot of \fig{UFHwithri}. In the $\log$-$\log$ plot, we see that the ratio $F_{\mathrm{static}}/F_{\mathrm{in-fall}}$ is unity initially. It then decreases with decreasing radius since $F_{\mathrm{in-fall}}$ is greater than $F_{\mathrm{static}}$ till the near- horizon region, where it switches trend diverging at the horizon subsequently due to non-regularity of the static flux. 

\section{Summary and Discussion} 
\label{sec:conclusion}

The thin null-shell gravitational collapse scenario can be studied unambiguously using the globally defined null coordinates defined in \cite{ms2013}. This is in contrast with the Kruskal extension of Schwarzschild metric where different vacuum states are defined for different boundary conditions to mimic the collapse. In that case, we have viz., (i) the Kruskal or the Hartle-Hawking vacuum state which is defined for the eternal black hole with respect to the global Kruskal coordinates and is symmetric under time-reversal thus describing a steady-state thermal equilibrium between the black hole and its surroundings, (ii) the Boulware vacuum state defined with respect to the Schwarzschild time and which reduces to the Minkowski vacuum state at large distance from the black hole thus having no particles from the point of view of the far away observer (similar to the Rindler vacuum) and (iii) the Unruh vacuum which is constructed to be time-asymmetric to reproduce the effects of a collapsing body yielding a time-asymmetric thermal flux from the black hole rather than a thermal bath. In this picture, however, with the globally defined null coordinates, the vacuum is the uniquely defined ``in"-vacuum in the past on $\mathcal{J}^-$ for the scalar field which we feel is more advantageous. The authors in reference \cite{ms2013} use the adiabatic expansion of detector response yielding the concept of an ``effective" temperature to study the Hawking phenomenon for various geodesic observers and also give the flux observed by an in-falling observer near the horizon. Building up on those lines we have considered the two local invariant quantities - energy densities and fluxes constructed from the renormalized stress-energy tensor and the $4$-velocity of an observer and its normal. The dependence on kinematics gives rise to very different perceptions of these two quantities given the parameters characterizing each trajectory in the spacetime. These quantities have been studied in the literature (for example see \cite{ford1993} and Appendix D of \cite{brout1995}) for the case of either an asymptotic observer who perceives the standard Hawking flux or for a radial in-faller from infinity and that too in the Unruh vacuum construction. \emph{It is important to note that no account of what is measured by the non-asymptotic observers other than these has been accounted for before with regard to the more natural in-vacuum state}.

For instance, for the static observers indicated by the their radius, the expressions for energy density and flux show quite some changes as the null shell collapses. Initially observed energy density and flux is zero with observer being in the Minkowski interior. It then shows a discontinuity when the null shell arrives which is quite understandable and then grows steadily entering the transient phase given by the thermalizing time which is different for different radii and reaches a saturated value which corresponds to the standard Tolman-shifted result in the Unruh vacuum for the static observers. The energy density for static observers also shows another feature that there is a negative energy region around near the horizon for which we have negative ingoing flux. Further the energy density diverges to the negative infinity at the horizon for the static observers while the flux diverges to the positive infinity. This is quite well known and related to the the infinite acceleration of the static observers at the horizon.  

We have next considered the radially in-falling observers marked by the energy per unit rest mass or the initial radius and the proper time at which the in-fall begins. We have explicitly calculated the energy density and flux as measured by this radially in-falling observer along its trajectory. It is observed that both energy density and flux are positive for these observers for larger initial radius matching initially with the static trajectory from which the in-falling observer diverts. But for an in-fall starting near the horizon the energy density can be negative to begin with. Both the observables are regular and finite for all in-falling observers except if the in-fall begins at the horizon. This is the case since any observer before in-falling would be thermalized to the static results and we see that the thermalizing time near the horizon is very quick. So except this case, our answer to the question, \emph{``What do the in-falling observers measure at the horizon?", is that they measure finite energy density and flux}. However, the ratio of flux seen by an observer falling from far off to that of the static observers is extremely small near the horizon. Both the quantities -- energy density and flux -- diverge at the singularity as the in-fall progresses inside the horizon. This feature of energy density and flux near the singularity is important for considering back-reaction on the geometry and is a work in progress which we shall address separately. 

\section*{Acknowledgments}

S.S and S.C. are supported by SPM grant from CSIR, India and would like to thank Prof. T. Padmanabhan for suggesting the project, helpful discussions and comments on the manuscript. We would also like to thank Kinjalk Lochan and Krishnamohan Parattu for reading the manuscript and giving useful comments. We also thank the referee for the useful comments.

\appendix
\section{Complete expressions for some quantities}
\label{app:A}

Below we give the complete expressions for the energy density and flux for a radially in-falling observer with $W = W(\delta e^{\delta-\kappa v^+})$:
\begin{align}
F= &\frac{\pi T_{H}^{2}}{12}\left[-4(1+4\delta)E\sqrt{E^{2}-\f{\delta}
{1+\delta}}\right.\nn\\
&\hspace{-10pt}\left.+\frac{(1+4W)(1+\delta)^{4}}{(1+W)^{4}}
\left(E+\sqrt{E^{2}-\f{\delta}{1+\delta}}\right) \right]\frac{1}{\delta
^{2}(1+\delta)^{2}}
\end{align}
\begin{align}
U=&\frac{\kappa
^{2}}{48\pi}\left[\frac{1+4W}{(1+W)^{4}}\frac{(1+\delta)^{2}}{\delta
^{2}}\left(2E^{2}+2E\sqrt{E^{2}-\f{\delta}{1+\delta}}\right.\right.\nn\\
&\left.\left. -\f{\delta}{1+\delta} \right)-\frac{4E^{2}(1+4\delta)}{\delta
^{2}(1+\delta)^{2}}+\frac{2(1+8\delta)}{\delta
(1+\delta)^{3}}\right].
\end{align}

\bibliographystyle{utcaps}
\providecommand{\href}[2]{#2}\begingroup\raggedright
\endgroup

\end{document}